\documentstyle[12pt]{article}
\begin{document}
\newcommand{\3}{\ss}
\newcommand{\dis}{\displaystyle}

\makebox[14cm][r]{PITHA 94/15}\par
\makebox[14cm][r]{TTP 94--03} \par
\makebox[14cm][r]{March 1994}\par
\vspace{.7cm}
\centerline{\Large A Transient New Coherent Condition of Matter:}
\par
\centerline{\Large The Signal for New Physics in Hadronic
Diffractive Scattering}
\vspace{1.cm}
\centerline{Saul Barshay$^{a)} $ and Patrick Heiliger$^{b)} $ }
\par
\centerline{$^{a)} $III. Physikalisches Institut (A), RWTH Aachen}
\par
\centerline{D--52056 Aachen, Germany}
\centerline{$^{b)} $Institut f\"ur Theoretische Teilchenphysik} \par
\centerline{Universit\"at Karlsruhe}\par
\centerline{D--76128 Karlsruhe, Germany}\par
\normalsize
\vspace{1.cm}

\begin{abstract}
We demonstrate the existence of an anomalous structure in the data on
the diffractive elastic scattering of hadrons at high energies and
small momentum transfer. We analyze five sets of experimental data
on $p(\overline{p})-p $ scattering from five different experiments
with colliding beams, ranging from the first-- and second--generation
experiments at $\sqrt{s} = 53 $ GeV to the most recent experiments
at 546 GeV and at 1800 GeV. All of the data sets exhibit a localized
anomalous structure in momentum transfer. We represent the anomalous
behavior by a phenomenological formula. This is based upon the idea
that a transient coherent condition of matter occurs in some of the
intermediate inelastic states which give rise, via unitarity, to
diffractive elastic scattering. The Fourier--Bessel transform into
momentum--transfer space of a spatial oscillatory behavior of matter
in the impact--parameter plane results in a small piece of the
diffractive amplitude which exhibits a localized anomalous behavior
near a definite value of $-t $. We emphasize in addition, possible
signals coming directly from such a new condition of matter, that
may be present in current experiments on inelastic processes.
\end{abstract}

\newpage
There exists an anomalous structure in the diffractive elastic
scattering of hadrons at high energies and small momentum transfer.
This anomaly exists in at least five sets of experimental data from
five different experiments on $p(\overline{p})-p $ scattering. These
experiments have been performed with colliding beams over a period
of some twenty years, at center--of--mass energies, $\sqrt{s} $,
equal to 53 GeV \cite{barbi}, \cite{schub}, \cite{amal}, 546 GeV
\cite{bozzo}, \cite{aug} and 1800 GeV \cite{amos}. The anomalous
behavior in the differential cross section consists of a small,
narrow peak at a point near to a four--momentum transfer $-t
\simeq 0.08 \, (\mbox{GeV/c})^2 $. This is followed by a marked dip
near $-t \simeq 0.10 \, (\mbox{GeV/c})^2 $ and by a subsequent
narrow peak at $-t $ in the range $0.12 - 0.16 \, (\mbox{GeV/c})^2 $.
The anomalous behavior can be exhibited by taking the ratio of the
experimental data to a good--fit, smoothly--varying theoretical or
phenomenological representation of all of the data at small momentum
transfers. It can also be exhibited in absolute value by taking
the difference between the experimental data and the
smoothly--varying theoretical representation. This is seen in Fig. 1
which, for illustration, we reproduce from an early analysis
\cite{osc} of the apparent anomaly in the data from the
first--generation collider experiment at 53 GeV \cite{barbi},
\cite{white}. In this paper we analyze the new $\overline{p} p $
data \cite{aug} at 541 GeV from the UA4/2 experiment at CERN and the
$\overline{p}p $ data \cite{amos} at 1800 GeV from the E--710
experiment at Fermilab. We also analyze the data \cite{bozzo} at 546
GeV from the original UA4 experiment. We go back and analyze the
very extensive $pp $ data at 53 GeV from the second--generation
experiment at the CERN intersecting storage rings. We find that a
localized anomalous behavior is present in all of these sets of
data. It also appears to be present in the fixed--target experiments
performed at Fermilab with very high statistics \cite{schiz} on
$\pi^{\mp} p $ elastic scattering at 200 GeV/c, (commented upon
further below). On the basis of the present analysis of the data, we
consider that there is evidence for an unusual phenomenon which is
giving a signal in hadronic diffractive scattering. This is probably
new physics. Guided by a definite physical possibility \cite{osc},
we represent the anomalous behavior in the data by a
phenomenological formula. The idea is that in the physical
intermediate states which occur from the dominant inelastic
processes that give rise, via unitarity, to diffractive elastic
scattering, there is a small fraction involving the formation of a
coherent condition in hadronic matter. This system is relatively
long--lived due to small effective velocities for signal
propagation within it. The system becomes extended over the
impact--parameter plane. In this plane, there occurs within the
system a coherent oscillatory behavior of matter, characterized by a
definite non--zero momentum. The Fourier--Bessel transform of this
spatial oscillatory behavior into momentum--transfer space gives
rise to a small piece of the diffractive amplitude which exhibits a
localized anomalous behavior near a definite value of $-t $. An
oscillatory behavior in $t $ occurs in the region near this value.
This is because of the approximate localization of the center of
mass of the coherent system which moves to a distant point along the
collision axis before the system breaks down into pions. In
concluding we emphasize in addition possible signals coming directly
from such a coherent system, signals which may be present in past
and current experiments on inelastic processes \cite{chli},
\cite{bott}, \cite{soph}. \par
\newpage
To augment Fig. 1, we begin our present analysis by exhibiting in
Fig. 2 the ratio of the experimental differential cross section
$\dis {d\sigma_{el} (pp) \over dt} $ at 53 GeV, from the
second--generation ISR experiment \cite{schub}, \cite{amal} to a
good theoretical representation \cite{gold}, \cite{rein} (on a
$\chi^2 $--basis $[ $F1$] $) of all of the data from $-t \simeq 0 $
to $-t \simeq 2 \, (\mbox{GeV/c})^2 $. In Fig. 2 as in Fig.1, we are
concentrating our analysis on the region of a hypothetical, quite
localized anomalous behavior, that is the region $0.07 < -t < 0.2 \,
(\mbox{GeV/c})^2 $. In Fig. 2 one observes the anomalous structure,
the same as that in Fig. 1.  \par
\smallskip
With a big jump in energy, we turn to the new UA4/2 data \cite{aug}
for $\dis {d\sigma_{el}(\overline{p}p) \over dt} $ at 541 GeV. In
Fig. 3 we show the ratio of this data to a good theoretical
representation \cite{gold}, \cite{rein}, \cite{own1} (again on a
$\chi^2 $--basis) of all of the data $[ $F2$] $. The data from the
UA4/2 experiment stops at the last interval $[ $F3$] $ $0.117 \leq
-t \leq 0.12 \; (\mbox{GeV/c})^2 $. We promptly augment this data
with the data from the first UA4 experiment \cite{bozzo} at 546 GeV;
the ratio is shown in Fig. 4 for the relevant region $0.07 < -t <
0.2 \, (\mbox{GeV/c})^2 $. Aspects of the anomalous structure at 53
GeV appear again in the data from the two separate experiments at 541
and 546 GeV. The evident part of the structure at 53 GeV, the peak
at $-t \simeq 0.16 \, (\mbox{GeV/c})^2 $, has moved in to about
$0.12 \, (\mbox{GeV/c})^2 $. The dip at $-t \simeq 0.12 \,
(\mbox{GeV/c})^2 $ has moved in to about $0.10 \, (\mbox{GeV/c})^2
$, and there is again a lesser peak at about $0.08 \, (\mbox{GeV/c})
^2 $. In the more extensive data from the first UA4 experiment, there
may be a dip at about $0.14 \, (\mbox{GeV/c})^2 $, followed by a
lesser peak as $-t $ moves out toward $0.2 \, (\mbox{GeV/c})^2 $.
Actually, the indication of structure in the data is directly
visible in the data for $\dis {d\sigma_{el} \over dt} $ near $-t =
0.1 \, (\mbox{GeV/c})^2 $: observe the insert in Fig. 2 of ref.
\cite{gold}. \par
\smallskip
We come now to the data \cite{amos} for $\dis {d\sigma_{el}
(p\overline{p}) \over dt} $ at $\sqrt{s} = 1800 $ GeV, the highest
energy presently measured. The ratio of this data to a theoretical
representation \cite{own1} is shown in Fig. 5. The structure at
$-t \simeq 0.12 \, (\mbox{GeV/c})^2 $ is again present, as is the
possible structure following this. Data points in the region of the
dip at $-t \simeq 0.10 \, (\mbox{GeV/c})^2 $ are lacking in this
experiment. The CDF collaboration at Fermilab has measured
$\dis d\sigma_{el} \over dt $ for $\overline{p}p $ scattering at
546 and 1800 GeV \cite{cdf}. The data is not tabulated in the paper
\cite{cdf}; at 546 GeV, data points are plotted in Fig. 9a of the
paper only out to $-t \le 0.08 \, (\mbox{GeV/c})^2 $. At 1800 GeV
the data for $\dis {d\sigma_{el} \over dt} $, measured in each
spectrometer arm separately, indicates visually in Fig. 9c of the
paper the possibility of structure near $-t = 0.1 \, (\mbox{GeV/c})
^2 $. This is so in particular for the data measured in arm --0
which detected symmetrically scattered elastic events with respect
to the beam. The new experiment on elastic scattering being
performed at Fermilab should try to probe the region $0.06 \le -t <
0.2 \, (\mbox{GeV/c})^2 $ with new precision. \par
\bigskip
Before going on to a more detailed analysis of the anomalous
structure, we emphasize an essential point. The evidence for a local
structure in the diffraction pattern does not depend upon the
theoretical details of the smoothly--varying representation of
all of the data which is used as a reference for comparison (to
form the above ratios). As long as the theory represents the data
well overall at any of the above energies, the local structure will
appear in the ratio. As evidence of this fact, we note that in the
original analysis \cite{osc} of the first 53 GeV data \cite{barbi}
reproduced in Fig. 1, the comparison was made to an adequate early
phenomenological representation \cite{white} based upon
parameterization of the eikonal \cite{glaub}, \cite{chou},
\cite{dur}. Indeed, the deviations are directly visible in the data
for $\dis {d\sigma_{el} \over dt} $ itself: note Figs. 2c and 2d
of reference \cite{barbi}. Recently, the new data \cite{aug} at 541
GeV has been compared to a further good representation \cite{wu1}
based upon the eikonal \cite{glaub}, \cite{chou}, \cite{dur} with
emphasis on the $t $--region of Coulomb--nuclear interference
\cite{wu2}. The authors have noted \cite{wu2} that there appears to
be oscillatory structure in the last bins of $t $ to which the data
extends. We shall have additional evidence below where we comment
upon comparison of the fixed--target $\pi^{\mp}-p $ data \cite{schiz}
from Fermilab with a smoothly--varying, phenomenological
representation \cite{schiz} of all of that data. \par
\smallskip
In our present analysis we have used a form for the $p(\overline{p})-
p $ diffractive (imaginary) amplitude which contains a new general
structure; this structure is the result of a derivation \cite{own2},
\cite{gold}, \cite{rein}, \cite{own1} which incorporates fluctuations
\cite{gold} in the eikonal into the general geometric picture
\cite{glaub}, \cite{chou}, \cite{dur}. Since we have published
comprehensive calculations \cite{gold}, \cite{rein}, \cite{own1} of
the experimental consequences of the new structure for energies up
to 1000 TeV, as well as detailed comparison \cite{rein} with another
good representation \cite{wu1} of the present data, we make only a
couple of general statements here. One is that at super--high
energies which we have estimated \cite{own1} to be $\sqrt{s} \sim
1000 $ TeV, the new structure closely approaches a new limiting form
\cite{own2} for the $S $--matrix in impact--parameter $(b) $ space
as a function of energy $S (b,s ) \sim 1/ (1+ s^{\lambda} f(b)) $
where $\lambda (s) $ approaches a small limiting value of the order
of magnitude of 0.15. The second remark is that there is a very
important physical difference between the behavior of the new
structure \cite{own2} in the geometric picture and the behavior of
the usual structure \cite{glaub}, \cite{chou}, \cite{dur} with
parameterization of the eikonal \cite{wu1}, as $\sqrt{s} $ increases
into the TeV range. The new structure gives rise to more increase
\cite{own1} in the total cross section in each succesively higher
domain of $\sqrt{s} $. Via the dispersion relation \cite{ugo},
\cite{own1}, this gives rise to a slowly but steadily increasing
value of $\rho $, the ratio of real to imaginary part of the
hadronic amplitude. We predict that $\rho $ rises to a value of about
\cite{own1} $17 \; \% $ at 17 TeV instead of decreasing to less
\cite{wu1} than $12 \; \% $. In summary, with $Im \; \overline{F}
(s,t) $ given explicitly by eqs. (1, 2, 3) in the second paper of
reference \cite{own1}, our theoretical representation is
\begin{equation}
{d\sigma \over dt} = \left |i \: Im \, \overline{F}(s,t) + \rho
{d\over dt} \left \{ t \: Im \, \overline{F}(s,t) \right \} \pm
{2\alpha \over |t| } \, e^{\mp i \alpha \phi(t)} \, G^2 (t) \right |
^2
\end{equation}
Although it becomes negligibly small in the region of $t $ that we
are emphasizing in this analysis of anomalous structure, we have
included the Coulomb amplitude \cite{ugo}, \cite{aug} with $\alpha
= 1/137$, $\phi(t) = \{ \ln ({0.08 \over |t|}) - 0.577 \} $, $G(t)
= 1/(1 + {|t| \over 0.71} )^2 $, and the $+ (-) $ sign relevant for
$\overline{p} (p)-p $ scattering. We include the contribution to
$\dis {d\sigma \over dt} $ of the real part of the hadronic
amplitude, proportional to $\rho $, using Martin's approximate form
\cite{mart} with empirical values for $\rho $. For $pp $ scattering
\cite{ugo} at 53 GeV $\rho = 8 \, \% $; for $\overline{p}p $
scattering we use our calculated \cite{own1} values $\rho = 13.9 \,
\% $ at 546 GeV and $\rho = 15 \, \% $ at 1800 GeV. These
predictions agree with experiment \cite{aug}, \cite{710}. We have
checked that including the real amplitude in an alternative manner,
simply in the form $\rho \; Im \, \overline{F} $, a common
phenomenological practice \cite{aug}, \cite{schiz}, \cite{wu1},
\cite{ugo}, \cite{710}, does not influence the actual evidence for
the anomalous structure. We comment upon this in detail below. \par
\bigskip
We proceed now to extract the anomalous structure as a definite
quantity. The addition of a small unusual diffractive amplitude
$\Delta (s, t) $ to the main diffractive amplitude $\overline{F}
(s,t) $ results in an addition to the differential cross section of
a small piece given by
\begin{equation}
\Delta \left ({d\sigma \over dt} \right ) = 2 \overline{F}(s,t)
\Delta (s,t)
\end{equation}
A good way to extract the anomalous amplitude $\Delta (s,t) $ (a
quantity measured in $\{\mbox{mb}/(\mbox{GeV/c})^2\}^{1/2} $ ) is
then to study the quantity $D $ defined by
\begin{eqnarray}
D(s,t) & = & { \left \{ \left ({d\sigma \over dt} \right)_{expt.}
- \left({d\sigma \over dt} \right ) \right \} \over \left ( {d\sigma
\over dt }\right )^{1/2} } \nonumber \\
& = & \left ({d\sigma \over dt} \right )^{1/2} \left \{R(s,t) - 1
\right \}
\end{eqnarray}
where $\dis {d\sigma \over dt} $ is the theoretical representation
in eq. (1) and $R(s,t) $ are the ratios exhibited in Figs. (2--5).
This we have done with the results shown in Fig. 6 for the
second--generation $[ $F4$] $ $pp $ data at 53 GeV, in Fig. 7 for
the new UA4/2 $\overline{p}p $ data at 541 GeV, in Fig. 8 for the
original UA4 data at 546 GeV, and in Fig. 9 for the E--710
$\overline{p}p $ data at 1800 GeV. In Fig. 10 we have placed the
data from 546 and 1800 GeV on a single graph in order to clearly
exhibit the consistent pattern from these two different scattering
experiments at the highest present energies. The similar definite
structure in Figs. (6--10) constitutes evidence for a localized
anomalous amplitude. A prominent aspect in this amplitude is a small
narrow peak which is at $-t \simeq 0.156 \, (\mbox{GeV/c})^2 $ at 53
GeV; this peak moves in to $-t \simeq 0.12 \, (\mbox{GeV/c})^2 $ at
546 and 1800 GeV. There is a dip near $-t = 0.1 \, (\mbox{GeV/c})^2
$ and a peak near $-t = 0.08 \, (\mbox{GeV/c})^2 $. There is a
damped, slightly irregular oscillatory behavior prior to this peak
at 53 GeV. The oscillatory behavior, present on both sides of the
peak, becomes more pronounced, more rapid and more regular at 546
GeV. With a magnitude of about $0.35 \; \{\mbox{mb}/(\mbox{GeV/c})^2
\}^{1/2} $ the peak amplitude (times 2) is $10 \, \% $ of $\dis
({d\sigma \over dt} )_{expt.}^{1/2} $ at $-t \simeq 0.156 \,
(\mbox{GeV/c})^2 $ at 53 GeV. At 546 GeV the anomaly is about
$6.5 \, \% $ of $\dis ({d\sigma \over dt}  )_{expt.}^{1/2} $
at $-t \simeq 0.12 \, (\mbox{GeV/c})^2 $; the reduction factor is
simply the ratio of total cross sections at 53 and 546 GeV, i.e.
$\sim { (}{42 \, mb \over 61 \, mb } { )} $. The
uncertainty in our extraction of the anomalous amplitude as a
definite quantity lies essentially in the uncertainty in overall
scale shifts of the ratios $R $ in Figs. (2--5). Overall scale
shifts do \underline{not} affect the evidence for a similar localized
structure in all of these ratios. They do affect the extraction of
a consistent anomalous amplitude. At 53 GeV and 546 GeV the $\dis
({d\sigma \over dt})_{expt.} $ have an uncertainty of about $2 \,
\% $, corresponding to about $\pm 1 $ mb in the total cross sections.
Our theoretical representation in eq. (1) has at least a similar
uncertainty. Thus the overall scale of $R $ in Fig. 2 and in Fig. 4
has a $4-5 \, \% $ uncertainty $[ $F5$] $. In Fig. 2 we have already
placed the $R $ essentially uniformly down by lowering $\dis (
{d\sigma \over dt})_{expt.} $ by $2 \, \% $ and raising $\dis
{d\sigma \over dt} $ by $3 \, \% $. This 53 GeV data is in
agreement with the 53 GeV data from the different experiment shown
in Fig. 1. In Fig. 4 for the $R $ at 546 GeV we have lowered
$\dis ({d\sigma \over dt})_{expt.} $ by $2 \, \% $ and raised
$\dis {d\sigma \over dt} $ by $2 \, \% $. At 1800 GeV, where our
$\dis {d\sigma\over dt } $ constitutes a prediction \cite{own1}
based upon two parameters $[ $F6$] $ which are fit to the total
cross sections at 53 and 546 GeV, we have raised $R $ in Fig. 5
uniformly by raising $\dis ({d\sigma \over dt})_{expt.} $ by $5 \,
\% $. This is because we believe that the central value of the
total cross section ($\sigma_t = 72.8 \pm 3.1 $ mb) obtained in
this experiment \cite{amos} is low. We predict \cite{own1} about 76
mb; it is noteworthy that the CDF collaboration at Fermilab has
recently stated \cite{cdf2} a value of $80.03 \pm  2.24 $ mb. The
size and the overall structure of the extracted anomalous amplitude
is really consistent from 53 up to 1800 GeV. Also the inclusion of
the real hadronic amplitude in the alternative manner as $\rho \;
Im \, \overline{F} $ in eq. (1) results in a consistent structure.
The effect at 53 GeV is negligible since $\rho^2 \approx 0.64 \, \% $
empirically \cite{ugo}. The presence of the anomalous structure
at 53 GeV is thus the best evidence for its not depending
essentially upon $\rho $. The quantitative effect at 541--546 GeV
($\rho^2 \approx 2 \, \% $) is essentially an overall rescaling of
the extracted anomalous amplitude: The peaks are a little lower and
the dip correspondingly a little deeper, by about 0.1 $\{\mbox{mb}/
\mbox{(GeV/c)}^2 \}^{1/2} $. This effect is immediately
understandable, because Martin's theoretically--motivated form for
taking account of the $\rho $ term in eq. (1) has a smooth shallow
minimum (zero) in the region of $-t $ between 0.1 and 0.2 $(\mbox
{GeV/c})^2 $; this is absent in the alternative form. This slight
overall scale change can be compensated by an overall rescaling of
the ratios $R $ in Figs. (3, 4), back up by about $2 \, \% $. The
\underline{structure} in the anomalous amplitude is not changed,
although its precise size as we have extracted it, has some
uncertainty related to the correlated uncertainties in the overall
scale of the $R $ and in the manner of taking account of the small
contribution to $\dis {d\sigma \over dt} $ from the $\rho ^2 $
term. We remark that we have generated and studied some 200 plots to
check out these statements. Thus we now attempt to quantitatively
describe the anomalous structure in terms of the properties of a
physical model. \par
\bigskip
{}From the outset we have been motivated in our analysis to
consistently isolate the localized anomaly in all of the data by a
definite physical model whose \underline{general} characteristics
would give rise to such behavior. Diffractive elastic scattering is
the ``shadow'' of all inelastic processes. In a small fraction of
these intermediate states the formation of a hypothetical coherent
condition in a system of hadronic matter evolves. The system in
which this coherent condition develops is hypothetically long--lived
on the usual scale of hadronization times ($\sim 2 \cdot 10^{-23} $
sec), and the system becomes extended over the impact--parameter
plane (the plane perpendicular to the collision axis). A coherent
oscillatory behavior of matter develops within the system in the
plane. The coherent oscillation is characterized by a fairly
definite non--zero momentum. The relevant Fourier--Bessel transform
\cite{glaub} of this spatial oscillatory behavior into
momentum--transfer space (the space of transverse momentum at small
scattering angles with $\sqrt{-t} \simeq {\sqrt{s} \over 2} \Theta $
) gives rise, in general, to a small piece of the diffractive
amplitude which exhibits a localized anomalous behavior near a
definite value of momentum transfer given by the non--zero momentum
$\sqrt{-t_0} $. This is the essential feature of the physical idea
and we parameterize this by a factor in the anomalous amplitude,
$e^{-R_b^2 (\sqrt{-t} - \sqrt{-t_0})^2/4} $. It is possible for the
center of mass of the coherent system to move rather rapidly along
the collision axis in the collision c.m., and thus the system's
dimension is Lorentz contracted in the longitudinal direction.
During its extended intrinsic lifetime prior to its breaking down
into pions (which lifetime is Lorentz--dilated by the $\gamma (s)
$ factor from its motion), the c.m. of the coherent system can reach
a fairly definite, distant point (from the collision origin) along
the collision axis. The ordinary Fourier transform of an
approximately localized position for the system to break down, into
the space of a relevant conjugate momentum which is the change in
the longitudinal momentum $ \sim -t /\sqrt{s} $ in the diffractive
scattering, gives rise in general to an oscillatory piece of the
amplitude. We parameterize this secondary feature by an oscillatory
factor in the anomalous amplitude which is either damped and a bit
irregular as described by a Bessel function $J_0 $, or is not damped
and regular as described by a cosine. The general simplified
structure of the anomalous amplitude, with possible energy dependence
of the parameters, is thus
\begin{equation}
2 \, \Delta (s,t) = A(s) \; e^{-R_b^2(s)(y-y_0(s))^2/4} \, \cdot
\cases { \;\;\;\;\;\;
{J_0 \left \{ \left ({T\over 2 m(s)} \right ) \left (y^2 - y_0^2(s)
\right ) \right \} } &  \cr
{\mbox{or} \;  \cos \left \{ \left ({T\over 2 m(s)} \right
) \left (y^2 - y_0^2(s) \right ) \right \}  } &  \cr }
\end{equation}
\begin{eqnarray}
& &
\mbox{with} \;\;\; y = \sqrt{-t}, \;\;\;  \!\! A  \; \mbox{in} \;
\left \{ \mbox{mb/(GeV/c)}^2 \right \}^{1/2} \nonumber
\end{eqnarray}
Note that in the argument of the oscillating factor the $1/\sqrt{s} $
energy dependence coming from the longitudinal momentum--transfer
variable is cancelled by the $\sqrt{s} $ coming from time--dilation
factor $\gamma(s) = \sqrt{s} / 2m(s) $ where $m(s) $ is an
effective mass--like parameter which includes $1/\lambda $ where
$\lambda $ is the fraction of $\sqrt{s}/2 $ involved in the system's
motion ($0 < \lambda < 1$) $[ $F7$] $. The non--zero momentum
characterizing coherent oscillation in the impact--parameter plane
is $y_0 $. Apart from the overall amplitude $A(s) $ which depends
also upon details of the particular collision process, there are
intrinsic length ($R_b $) and time ($T $) parameters, with $R_b
(s) $ essentially determining the narrowness of the structure around
$y_0 $. Although evolving to some extension on the impact--parameter
plane significantly larger than the dimension characterizing the
initial collision $\sim 1/m_{\pi} = 1.4 $ fm, the condensate
dimension is finite and this limits attaining a very definite value
for $y_0 $, because of the uncertainty of order $1/R_b(s) $. The
hypothetical coherent system is evolving toward some equilibrium
condition, but it is limited by the intrinsic, though extended
hadronization time $\sim T $. The parameters $y_0, R_b $ and $T $
characterize the essential physical aspects of the coherent system,
and are necessary $[ $F7$] $. We have investigated a definite model
patterned after studies of pion condensation in a nuclear medium
\cite{brown}, but involving quasi--pions (correlated
quark--antiquark pairs with small $(\mbox{mass})^2 $) propagating
in a dense medium of ``dressing'' quarks and antiquarks, looking
for the onset of a coherent condition \cite{phot} that contributes
a negative energy density in the medium. The essential
\underline{dynamical} element which makes this possible is the
existence of a pseudoscalar (quasi--)boson with quite small $(\mbox
{mass})^2 $ and strong, low--energy, $P $--wave interaction with
fermions. This allows for the possibility of cancellation of the
positive, kinetic and $(\mbox{mass})^2 $ terms by the attractive
momentum--dependent interaction \cite{brown}. For orientation we
have attempted to roughly estimate \cite{phot}, in such a coherent
system, the parameters relevant for the anomalous amplitude in eq.
(4). One result \cite{brown} is that $T $ is significantly increased
from an apriori hadronization time of about $2 \cdot 10^{-23} $ sec
because the effective velocity for signal propagation by which
parts of the medium communicate (and hence hadronization is
carried out) is much smaller than that of light. This is related
to the fact that $R_b $ increases only as $\sqrt{T} $. The region of
the coherent momentum is of the order of magnitude of $2 \, m_{\pi}
$. Using estimated parameters \cite{phot} in eq. (4) $y_0 = 0.395 \,
(\mbox{GeV/c}), R_b = (2.7/m_{\pi}), \{T/m \} = \{22/m_{\pi} (1 \,
\mbox{GeV}) \} $ with the $J_0 $, and taking $A = 0.35 $, we obtain
the curve for the anomalous amplitude shown in Fig. 6. In the big
jump in energy from 53 to 546 GeV, one guesses that the coherent
system may become more ``sharply'' defined, that is $R_b(s) $ could
increase, the parameter $\left \{ {T\over m(s)} \right \} $ could
increase and the oscillatory factor can become regular with little
damping. In addition, the increasing density of the fermionic medium
in general allows for a somewhat smaller value of momentum $y_0 $.
The single curve compared to all of the data in Figs. 7, 8, 9, 10 is
from eq. (4) with the cosine and with $A = 0.35, \; y_0(s) = 0.346 \,
(\mbox{GeV/c}) $, $\left \{ T/m(s) \right \} $ increased by 2 to
$\{ 44/m_{\pi} (1\, \mbox{GeV}) \} $, and $R_b(s) $ increased by
$\sqrt{2} $ to $(3.8/m_{\pi}) $. \par
\bigskip
We have isolated a localized anomalous structure in five sets of
collider data on $p(\overline{p})-p $ diffractive scattering. This
anomaly apparently also exists in the high--statistics fixed--target
data from Fermilab \cite{schiz}, in particular for $\pi^{\mp}p $
scattering at 200 GeV/c. For $\pi^-p $ scattering this can be seen
directly in the differential cross section data near to $-t = 0.1 \,
(\mbox{GeV/c})^2 $ in Fig. 14 of the first paper in reference
\cite{schiz}. It can be seen well in Fig. 6 of the second paper,
which we have reproduced here in Fig. 11. There, the local slopes
$b(t) $ from the data $\dis ({d\sigma \over dt} \propto e^{b(t)t}
) $ are compared to those occurring in a smoothly--varying
phenomenological fit to all of the data established by the
experimentalists themselves \cite{schiz}. The anomalous, low value
of the local slope just below $-t = 0.1 \, (\mbox{GeV/c})^2 $ is
evident for $\pi^-p $ and $\pi^+p $, and is possibly also present
for $pp $. \par
\bigskip
We have shown that the localized anomalous structure in hadronic
diffractive scattering can be represented $[ $F8$] $ by certain
rather general features of a model based upon a transient new
coherent condition of matter occurring in some of the inelastic
intermediate states. Such a coherent condition could give direct
unusual signals in current (and past) experiments on inelastic
processes. An important signal would be an anomalous level
of soft photons. If charged currents exist during the evolution
of the coherent system, then it has been shown that soft
photons will be emitted \cite{saul}. In the system, the energies
can be very small on the usual scale of $m_{\pi } $, down to
$E_{\gamma } $ of the order of $1/T \simeq 6 $ MeV with a
bremsstrahlung--like spectrum \cite{saul}, \cite{phot}. We have
estimated \cite{phot} that the level of anomalous emission can reach
several times the level of the bremsstrahlung from incoming and
outgoing charged hadrons. Since the latter is at the level
\cite{soph} of about $2 \, \% $, this corresponds to a soft photon
occurring in of the order of $ 10 \, \% $ of inelastic collisions.
In fact this is about the same as the relative size of the anomalous
effect in diffractive scattering that we have analyzed in this
paper. Anomalous emission of soft photons has been observed
\cite{chli}, \cite{bott}, \cite{soph} and is being observed
presently \cite{soph2} in certain experiments at CERN. For a
coherent system without charged currents, an unusual signal would be
the materialization into some neutral pions with anomalously small
transverse momenta $\sim 1/R_b < 50 $ (MeV/c). These estimates for
very small momenta have used the parameters $T $ and $R_b $ from our
representation of the anomalous structure in Fig. 6.
\par
\bigskip
High--energy particle physics has been a science guided by careful
attention to the results of laboratory experiments (including
originally those using cosmic rays). Attention to all experimental
results, including unusual effects that do not seem to fit in,
is important today. In this spirit it would seem that the unusual
structure that we have exhibited in the present analysis of old
and new scattering data deserves further experimental investigation
$[ $F3$] $ in an effort to establish new physics. It can be a signal
for a transient new coherent condition of hadronic matter.

\newpage
\large
{\bf Added Note} \normalsize (April 19, 1994) \par
\bigskip
\noindent
The CDF collaboration has kindly sent us their (combined--arm)
data for $\large {d\sigma_{el} \over dt} $ at 1800 GeV. In Fig. 12
we show the ratio function for this data (stars). To facilitate
comparison, Fig. 12 also contains the ratio function for the E--710
data (circles) reproduced from Fig. 5. Both data sets are normalized
to $\sigma_{tot} (\mbox{theor}) \simeq 76 $ mb. We thank Paolo
Giromini and the CDF collaboration.

\newpage
\large
\noindent
{\bf Footnotes}
\normalsize
\begin{itemize}
\item[[F 1]] For the detailed demonstration of the $\chi^2 $ quality
             we refer the reader to Fig. 1 of reference \cite{rein}
             and Fig. 3 of reference \cite{gold}.
\item[[F 2]] Here we refer to the more complete data from the first
             UA4 experiment (reference \cite{bozzo}) in the domain
             $0 < -t < 1 \, (\mbox{GeV/c})^2 $. For the detailed
             demonstration of the $\chi^2 $ quality of the
             theoretical representation, we refer the reader to Fig.
             2 of reference \cite{gold} and Fig. 3 of reference
             \cite{rein}.
\item[[F 3]] S.B. thanks Maurice Hagenauer for explaining to him
             at CERN the technical reasons for this unfortunate fact.
             The UA4/2 experiment was of course designed to measure
             again the value of $\rho $ (reference \cite{aug}), and
             hence to concentrate upon the region of the main
             interference between hadronic and Coulomb real
             amplitude; this region is near $-t = 0.001 \,
             (\mbox{GeV/c})^2 $. Experiments to concentrate upon the
             limited $t $ region of the hypothetical anomalous
             structure have never been carried out. They should be.
\item[[F 4]] We mention here that we have examined the data at 62.5
             GeV (references \cite{schub}, \cite{amal}). There is
             also structure in this data as one approaches $-t =
             0.1 \, (\mbox{GeV/c})^2 $, but this data has a curious,
             complete absence of data points between $-t = 0.1 $ and
             $0.2 \, (\mbox{GeV/c})^2 $.
\item[[F 5]] The new but limited data at 541 GeV is not tabulated
             in $\{ \mbox{mb/(GeV/c)}^2 \} $ (reference \cite{aug}).
             We have fixed the ratio $R $ at the dip to agree with
             that at 546 GeV.
\item[[F 6]] These are the two parameters which determine the
             logarithmic growth with energy of the central opacity,
             given in eq. (4) of the first paper in reference
             \cite{own1}.
\item[[F 7]] The quanitity $\dis \{ {T\over m(s)} \} $ can of
             course be taken as a single phenomenological parameter,
             giving together with $A, \, y_0 $, and $R_b $, four
             physical parameters.
\item[[F 8]] We use the word ``representation'' because the main
             thrust of the present paper is to demonstrate and
             to isolate the anomalous structure in the data. The
             fact that the anomaly can be reasonably represented
             within a definite physical model perhaps lends some
             force to its existence.
\end{itemize}

\newpage
\large
\noindent
{\bf Figure Captions}
\normalsize
\begin{itemize}
\item[Fig. 1] The figure reproduced from ref. \cite{osc} shows in a)
              in the relevant region of $t $ the data on $\dis
              {d\sigma_{el} \over dt} $ from the first--generation
              CERN ISR experiment for $pp $ elastic scattering at
              $\sqrt {s} = 53 $ GeV \cite{barbi} in ratio to a
              smoothly--varying, good phenomenological fit from that
              time to all of the data \cite{white}. In b) the
              absolute difference between experiment and theory is
              shown. The curve is simply a phenomenological
              representation of the peak structure given in ref.
              \cite{osc}.
\item[Fig. 2] The data on $\dis {d\sigma_{el} \over dt} $ from the
              second--generation CERN ISR experiment for $pp $
              elastic scattering at $\sqrt{s} = 53 $ GeV
              \cite{schub}, \cite{amal} in ratio to a
              smoothly--varying, good theoretical representation of
              all of the data, using eq. (1) in the text below and
              eqs. (1, 2, 3) in the second paper of ref. \cite{own1}
              for $Im \, \overline{F}(s,t) $.
\item[Fig. 3] The data on $\dis {d\sigma_{el} \over dt} $ from the
              new UA4/2 experiment at the CERN $\overline{p}p $
              collider at $\sqrt{s} = 541 $ GeV \cite{aug} in
              ratio to a smoothly--varying, good theoretical
              representation of all of the data at 546 GeV
              \cite{bozzo}, using eq. (1) in the text below and
              eqs. (1, 2, 3) in the second paper of ref. \cite{own1}
              for $Im \, \overline{F}(s,t) $.
\item[Fig. 4] The data on $\large {d\sigma_{el} \over dt} $ from the
              UA4 experiment at the CERN $\overline{p}p $ collider
              at $\sqrt{s} = 546 $ GeV \cite{bozzo} in ratio
              to the same smoothly--varying, good theoretical
              representation as in Fig. 3.
\item[Fig. 5] The data on $\large {d\sigma_{el} \over dt} $ from the
              E--710 experiment at the Fermilab $\overline{p}p $
              collider at $\sqrt{s} = 1800 $ GeV \cite{amos}
              in ratio to a smoothly--varying, good theoretical
              representation of all of the data, using eq. (1) in the
              text below and eqs. (1, 2, 3) in the second paper of
              ref. \cite{own1} for $Im \, \overline{F}(s,t ) $.
\item[Fig. 6] The amplitude extracted from the data using
              eq. (3) and the ratio $R $ at 53 GeV from Fig. 2 with
              $\large {d\sigma \over dt} $ from eq. (1). The curve is
              the representation of the anomalous structure by eq.
              (4) with $A = 0.35, \, y_0 = 0.395 \, (\mbox{GeV/c}),
              \, R_b = (2.7/m_{\pi}) $, and $\{T/m \} = \{22/m_{\pi}
              (1 \, \mbox{GeV}) \} $ using the $J_0 $.
\item[Fig. 7] The anomalous amplitude extracted from the data using
              eq. (3) and the ratio $R $ at 541 GeV from Fig. 3 with
              $\large {d\sigma \over dt} $ from eq. (1). The curve is
              the representation of the anomalous structure by eq.
              (4) with $A = 0.35, \, y_0 = 0.346 \, (\mbox{GeV/c}),
              \, R_b = (3.8/m_{\pi}) $, and $\{T/m \} = \{44/m_{\pi}
              (1 \, \mbox{GeV}) \} $ using the cosine.
\item[Fig. 8] The anomalous amplitude extracted from the data using
              eq. (3) and the ratio $R $ at 546 GeV from Fig. 4 with
              $\large {d\sigma \over dt} $ from eq. (1). The curve
              is the same as in Fig. 7.
\item[Fig. 9] The anomalous amplitude extracted from the data using
              eq. (3) and the ratio $R $ at 1800 GeV from Fig. 5
              with $\large {d\sigma \over dt} $ from eq. (1). The
              curve is the same as in Figs. (7, 8).
\item[Fig. 10] The amplitudes at 546 and 1800 GeV from Figs. 8, 9 are
               placed on a single graph together with the
               representation of the anomalous structure by eq. (4).
\item[Fig. 11] The figure, a reproduction of Fig. 6 from the
               second paper in ref. \cite{schiz}, shows the data on
               the local slopes $b(t) $ $\large ( {d\sigma \over
               dt} \propto e^{b(t)t} ) $ for $\pi^{\mp } p $ and $pp
               $ elastic scattering from the Fermilab
               high--statistics, fixed--target experiments at 200
               GeV/c. The data is compared with the local slopes
               (solid curves) from good phenomenological fits to all
               of the data established by the experimentalists
               themselves (ref. \cite{schiz}). The dashed lines show
               the envelope of uncertainties from their overall fits.
\end{itemize}

\end{document}